\documentclass[conference]{IEEEtran}
\IEEEoverridecommandlockouts

\usepackage{cite}
\usepackage{amsmath,amssymb,amsfonts}
\usepackage{algorithmic}
\usepackage{graphicx}
\usepackage{subcaption}
\usepackage{textcomp}
\usepackage{xcolor}
\usepackage[hidelinks]{hyperref}
\def\BibTeX{{\rm B\kern-.05em{\sc i\kern-.025em b}\kern-.08em
    T\kern-.1667em\lower.7ex\hbox{E}\kern-.125emX}}
\begin{document}


\title{Noninvasive Anisotropic Identification of Magnetic Properties in Toroidal Shaped Magnetic Steel}

\author{\IEEEauthorblockN{ Xusen Qin}
\IEEEauthorblockA{\textit{Department of Electrical Engineering} \\
\textit{Eindhoven University of Technology}\\
Eindhoven, The Netherlands \\
x.qin@tue.nl}
\and
\IEEEauthorblockN{Reza Zeinali}
\IEEEauthorblockA{\textit{R\&D Department} \\
\textit{Kinetron - ASSAABLOY}\\
Tilburg, The Netherlands \\
reze.zeinali@assaabloy.com}
\and
\IEEEauthorblockN{Mitrofan Curti}
\IEEEauthorblockA{\textit{Department of Electrical Engineering} \\
\textit{Eindhoven University of Technology}\\
Eindhoven, The Netherlands \\
M.Curti@tue.nl }

}

\maketitle

\begin{abstract}
This research proposes a model-free inverse approach for identifying the nonlinear anisotropic single-valued magnetic constitutive characteristics of soft magnetic materials from boundary measurements. 
Using the harmonic approximation, the internal magnetic flux density $\mathbf{B}$ and magnetic field strength $\mathbf{H}$ are independently reconstructed from the measured boundary data, thus avoiding the limitation of using a predefined parametric B--H mapping function. Subsequently, the reconstructed fields are spatially averaged along selected collection lines to recover the B--H characteristics of the principal directions.
Numerical results demonstrate high reconstruction accuracy for unperturbed data and good robustness against Gaussian perturbations of the boundary measurements, with the overall mean error remaining below $5\%$ for a $10\%$ noise level. These results indicate that under the harmonic-field assumptions, the field-based reconstruction provides a simple and computationally efficient framework for non-invasive identification of nonlinear anisotropic magnetic constitutive characteristics.
\end{abstract}


\section{Introduction}

Accurate magnetic material data are essential for the design and numerical analysis of systems based on magnetic materials with magnetic cores from plain iron, stacked or soft magnetic composite. These systems range from magnetic lenses, actuators, motors and specialized equipment, based on soft magnetic composites, to name only a few.
In practice, the effective $B$--$H$ characteristics of a manufactured component often deviate from the nominal material data due to mechanical stress, cutting, assembly, and other manufacturing effects \cite{hofmannMagneticProperties2016}.
This motivates non-invasive identification methods that estimate the magnetic constitutive law directly from electromagnetic measurements performed on the actual component.

Inverse electromagnetic approaches have previously been employed to recover magnetic material characteristics from measured quantities in devices with non-uniform magnetic fields \cite{ioanExtractionBHRelation2002, abdallhMagneticMaterialIdentification2009}. 
Coupled experimental--numerical methods using global or local magnetic measurements have also demonstrated the feasibility of identifying the magnetic characteristics directly on an assembled electromagnetic device \cite{abdallhInverseApproach2010}. However, uhese methods may be limited by the chosen constitutive parameterization. Specifically, they typically yield a homogenized, global representation of the material properties, thereby failing to capture local variations. Furthermore, their accuracy and flexibility are inherently constrained by the predefined analytical functions or parametric models used to fit the measurement data.

In this work, we address the limitation of fitting measurements to specific predefined models through a model-free approach, pushing the characterization framework further to capture anisotropic behaviors. Specifically, we consider a toroidal soft magnetic core and aim to identify its unknown, single-valued magnetic constitutive characteristics directly from controlled current excitations and boundary magnetic measurements.
The measured quantities are formulated as so-called Cauchy data on the core boundary, consisting of the normal component of the magnetic flux density, $B_n$, and the tangential component of the magnetic field strength, $H_t$.
These boundary data are subsequently used as the input to the inverse problem for reconstructing the magnetic fields inside the core.

Instead of introducing a predefined parameterized constitutive model, this method reconstructs the internal magnetic fields $\mathbf{B}$ and $\mathbf{H}$ approximately independently under the harmonics-field assumption.
The reconstructed magnetic field is then sampled along selected acquisition lines, and a sequence of current excitations is used to generate magnetic operating points over a wide magnetic field range.
These operating points are finally employed to reconstruct the nonlinear $B$--$H$ characteristics in the principal material directions.
The robustness of the proposed identification procedure is further evaluated by introducing controlled Gaussian perturbations into the measured Cauchy data.
\section{Problem Setup}

\subsection{Identification of the Anisotropic Single-Valued Magnetic Constitutive Law}
In this study, the constitutive response is assumed to be separable along the principal material axes, with cross-component coupling neglected.
The anisotropic single-valued magnetic constitutive law is defined as a nonlinear vector mapping from the magnetic field strength $\mathbf{H}$ to the magnetic flux density $\mathbf{B}$,
\[
\mathbf{B} = \boldsymbol{\mathcal{B}}(\mathbf{H}),
\quad
\boldsymbol{\mathcal{B}} : \mathbb{R}^{d} \rightarrow \mathbb{R}^{d},
\]
where $d$ denotes the spatial dimension. For laminated electrical steel, the rolling direction (RD) and transverse direction (TD) are considered as the principal material directions. The corresponding single-valued magnetization characteristics are described by the one-dimensional constitutive relations:
\[
\begin{bmatrix}
B_{RD} \\
B_{TD}
\end{bmatrix}
=
\begin{bmatrix}
\mathcal{B}_{RD}(H_{RD}) \\
\mathcal{B}_{TD}(H_{TD})
\end{bmatrix}.
\]

These two curves characterize the nonlinear magnetic responses along the principal material axes and generally differ due to the material anisotropy \cite{Chwastek2013}. 
In this work, a model-free identification approach is adopted to construct the anisotropic constitutive mapping directly from the available magnetic measurement data, without assuming a predefined analytical form of the constitutive law. The identified mapping is then used to evaluate the magnetic flux density corresponding to an arbitrary magnetic field vector within the material plane.

\subsection{Toroidal Geometry and Measurement Setup}
\begin{figure*}[t]
    \centering
    \begin{subfigure}[t]{0.44\textwidth}
        \centering
        \includegraphics[width=\textwidth]{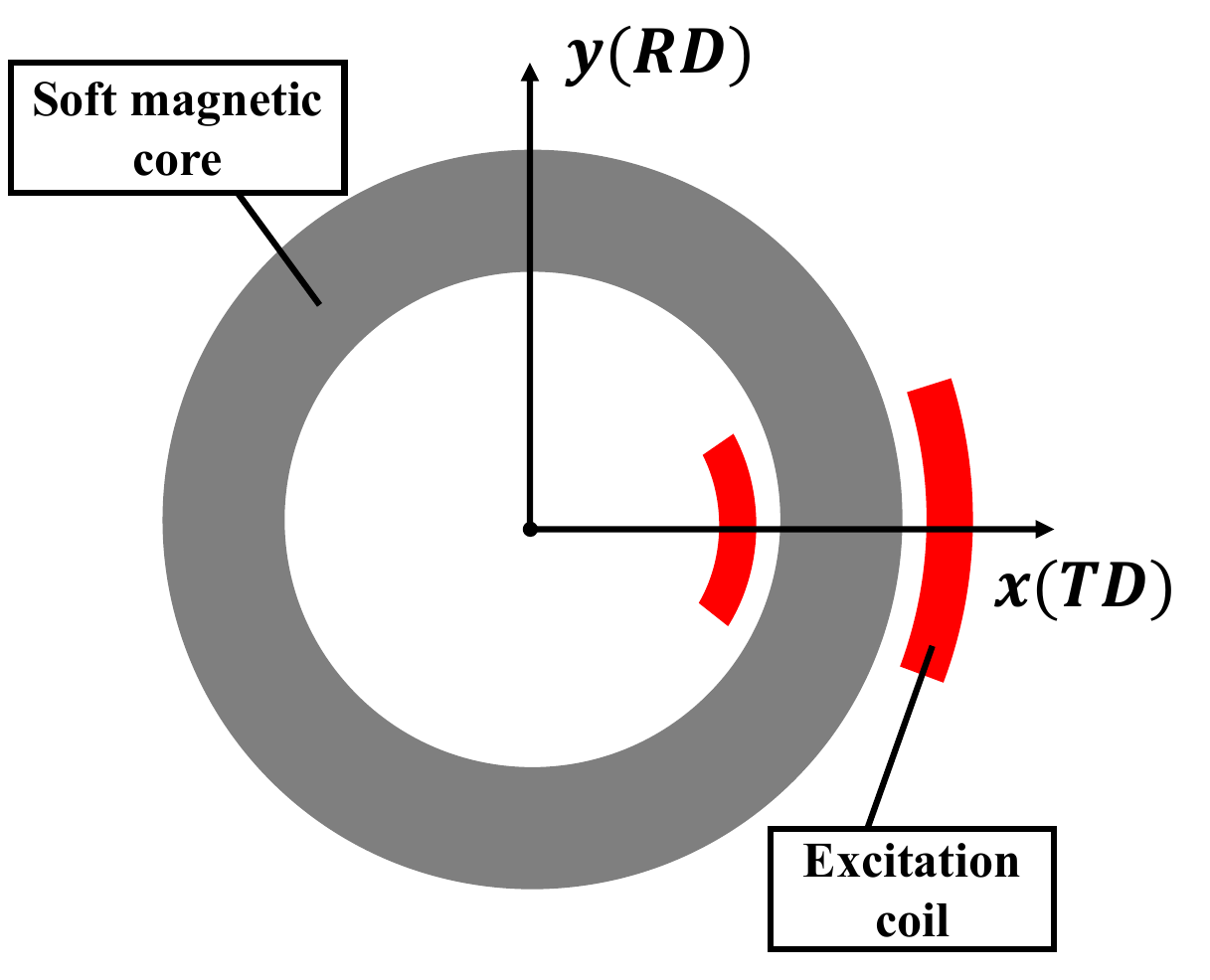}
        \caption{The 2D toroidal geometry.}
        \label{fig:topo}
    \end{subfigure}
    \hspace{0cm}
    \begin{subfigure}[t]{0.35\textwidth}
        \centering
        \includegraphics[width=\textwidth]{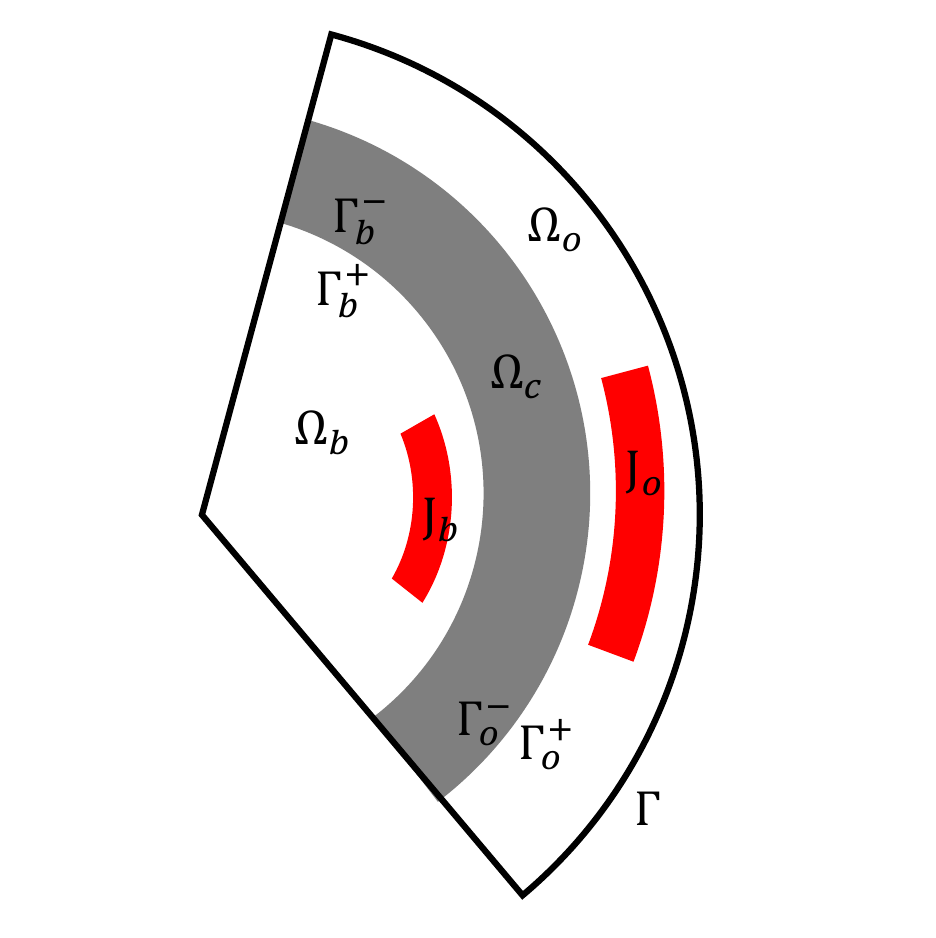}
        \caption{Computational domain.}
        \label{fig:topo_zoom}
    \end{subfigure}
    \caption{Computational topology for the proposed identification problem.}
    \label{fig:topology_combined}
\end{figure*}

Here, the soft magnetic core is assumed to be in-plane (2D) toroidal. This can be obtained for instance by having a sufficiently long toroid to produce negligible end effects. We define the principal material axes along the $x$-axis and $y$-axis, which correspond to the transverse direction (TD) and rolling direction (RD), respectively, together with the toroidal axis. To preserve symmetry, the excitation coil is placed symmetrically with respect to the $x$-axis (TD). The toroidal geometry is shown in Figure~\ref{fig:topology_combined}.

As described in Figure~\ref{fig:topo_zoom}, the computational domain $\Omega$ is decomposed as
\[
\overline{\Omega}
=
\overline{\Omega}_b
\cup
\overline{\Omega}_c
\cup
\overline{\Omega}_o,
\quad
\Omega_i \cap \Omega_j = \varnothing,
\quad i\neq j,
\]
where $\Omega_c$ represents the ferromagnetic core. The inner and outer
interfaces of the core are defined by
\[
\Gamma_b
=
\partial\Omega_b \cap \partial\Omega_c,
\quad
\Gamma_o
=
\partial\Omega_c \cap \partial\Omega_o,
\]
The superscripts \(-\) and \(+\) denote the radially inner and outer sides of each interface, respectively.
The outer boundary of the entire computational domain is denoted by $\Gamma$.

\section{Forward Problem Formulation}

\subsection{Forward Problem: Magnetostatic Model }

Under the magnetostatic assumption, the magnetic field in the two-dimensional domain is described by
\[
\mathbf{H} = \begin{bmatrix} H_x \\ H_y \end{bmatrix},
\quad
\mathbf{B} = \begin{bmatrix} B_x \\ B_y \end{bmatrix}.
\]
In the ferromagnetic core $\Omega_c$, the nonlinear anisotropic magnetic behavior is represented by the single-valued constitutive mapping

\begin{equation}
\mathbf{B} = \boldsymbol{\mathcal{B}}(\mathbf{H}) = \begin{bmatrix} \mathcal{B}_x(H_x) \\ \mathcal{B}_y(H_y) \end{bmatrix},
\quad
\text{in } \Omega_c.
\label{eq:BH-aniso}
\end{equation}
In the non-magnetic regions, the permeability is assumed to be equal to the vacuum permeability $\mu_0$.
Hence, the constitutive relation over the entire computational domain can be written as
\[
\mathbf{B} = \begin{cases}
\boldsymbol{\mathcal{B}}(\mathbf{H}), & \text{in } \Omega_c, \\[2mm]
\mu_0\mathbf{H}, & \text{in } \Omega\setminus\overline{\Omega}_c.
\end{cases}
\]

The magnetostatic problem is governed by
\begin{equation}
\begin{cases}
\nabla\cdot\mathbf{B} = 0, & \text{in } \Omega, \\[1mm]
\nabla\times\mathbf{H} = \mathbf{J}, & \text{in } \Omega,
\end{cases}
\label{eq:govern_eqs}
\end{equation}
with
\[
\mathbf{J} =
\begin{cases}
\mathbf{J}_b,
& \text{in } \Omega_b,\\[1mm]
\mathbf{0},
& \text{in } \Omega_c,\\[1mm]
\mathbf{J}_o,
& \text{in } \Omega_o.
\end{cases}\quad \int_{\Omega_b} \mathbf{J}_b \mathrm{d}S + \int_{\Omega_o} \mathbf{J}_o \mathrm{d}S =0
\]

The problem is solved together with the above piecewise constitutive relation.
Under the two-dimensional assumption, the magnetic vector potential $\mathbf{A}$ is written as,
\[
\mathbf{A}
=
\begin{bmatrix}
0\\
0\\
A_z(x,y)
\end{bmatrix}
=
A_z(x,y)\mathbf{e}_z,
\quad
\frac{\partial A_z(x, y)}{\partial z}=0,
\]
where $\mathbf{e}_z$ denotes the unit vector in the $z$-direction, the
magnetic flux density is expressed as
\[
\mathbf{B}
=
\nabla\times\mathbf{A}
=
\begin{bmatrix}
\dfrac{\partial A_z}{\partial y}\\[2mm]
-\dfrac{\partial A_z}{\partial x}\\[2mm]
0
\end{bmatrix}.
\]

A homogeneous Dirichlet boundary condition is imposed on the external boundary,
\begin{equation}
    \label{eq:DC_Gamma}
A_z = 0,
\quad
\text{on } \Gamma.
\end{equation}

\subsection{Boundary Measurements and Cauchy Data}
\label{section:measurement}

The required boundary quantities are the normal component of the magnetic flux
density and the tangential component of the magnetic field strength on the
interfaces.
Assuming that no surface current is present at the material interfaces, the normal component of $\mathbf{B}$ and the tangential component of $\mathbf{H}$ are continuous across each interface. Hence, for \(i\in\{b,o\}\)
\[
\mathbf{n}_i\cdot\mathbf{B}^{+} = \mathbf{n}_i\cdot\mathbf{B}^{-} = B_{n,i},
\quad
\mathbf{t}_i\cdot\mathbf{H}^{+} = \mathbf{t}_i\cdot\mathbf{H}^{-} = H_{t,i},
\quad
\text{on } \Gamma_i .
\]
Here $\mathbf{n}_i$ and $\mathbf{t}_i$ denote the unit normal and tangential vectors to $\Gamma_i$, respectively.

The available measurements form the Cauchy data on the two interfaces, denoted
for $i\in\{b,o\}$ as
\begin{equation}
\mathcal{C}
=
\left\{
\mathcal{C}_b,\,
\mathcal{C}_o
\right\},
\quad
\mathcal{C}_i
=
\left\{
B_{n,i},\,
H_{t,i}
\right\},
\quad \text{on } \Gamma_i.
\label{eq:Cauchy}
\end{equation}

In a possible experimental realization, these data can be obtained from the
air side using a calibrated three-axis Hall probe mounted on a non-magnetic
rotary and radial positioning system. At each excitation level, the probe scans
close to both interfaces and records the local vector $\mathbf{B}_{\mathrm{air}}$.
The required components are then determined from
\[
B_{n,i}=\mathbf{n}_i\cdot\mathbf{B}_{\mathrm{air}},
\qquad
H_{t,i}=\frac{\mathbf{t}_i\cdot\mathbf{B}_{\mathrm{air}}}{\mu_0}.
\]
The latter relation follows from $\mathbf{B}=\mu_0\mathbf{H}$ in air and the
continuity of tangential $\mathbf{H}$, allowing $H_{t,i}$ to be inferred without
placing a sensor inside the material.

In this numerical study, synthetic Cauchy data are extracted from the forward solution. The predefined B--H curves are used for forward data generation and error evaluation, from which boundary traces $B_{n,i}$ and $H_{t,i}$ are extracted as synthetic Cauchy data. The inverse reconstruction uses only these boundary traces and does not use the predefined constitutive laws.

\section{Inverse Problem Formulations}

\subsection{Harmonic Approximation and Field Reconstruction}
To reconstruct the magnetic fields inside the core without explicitly knowing the underlying nonlinear constitutive law, the magnetic field strength $\mathbf{H}$ and the magnetic flux density $\mathbf{B}$ are approximated independently by harmonic fields in $\Omega_c$. The reconstruction is driven by Cauchy data measured on the two interfaces $\Gamma_b$ and $\Gamma_o$. The harmonic approximation is defined by requiring the reconstructed fields to be both divergence-free and curl-free in the current-free core domain:
\begin{equation}
\begin{cases}
\nabla \cdot \widehat{\mathbf{H}} = 0,\\
\nabla \times \widehat{\mathbf{H}} = 0,
\end{cases}
\quad
\begin{cases}
\nabla \cdot \widehat{\mathbf{B}} = 0,\\
\nabla \times \widehat{\mathbf{B}} = 0,
\end{cases}
\quad
\text{in } \Omega_c.
\label{eq:inv_harmonic}   
\end{equation}

These harmonic fields are represented through the scalar components of potentials and
reconstructed from the corresponding Cauchy boundary data.
For the reconstruction of the magnetic field strength, a vector potential
$\boldsymbol{\psi}$ is introduced under the two-dimensional plane assumption as
\[
\boldsymbol{\psi}
=
\begin{bmatrix}
0\\
0\\
\psi_z(x,y)
\end{bmatrix},
\qquad
\widehat{\mathbf{H}}
=
\nabla\times\boldsymbol{\psi}
=
\begin{bmatrix}
\dfrac{\partial \psi_z}{\partial y}\\[2mm]
-\dfrac{\partial \psi_z}{\partial x}\\[2mm]
0
\end{bmatrix}.
\]

This representation automatically satisfies
\[
\nabla\cdot\widehat{\mathbf{H}}=0.
\]
Moreover, since no source current is present inside the magnetic core,
\[
\nabla\times\mathbf{H}=0
\quad
\text{in } \Omega_c.
\]
Therefore, the potential function satisfies the Laplace equation
\[
\Delta\psi_z=0
\quad
\text{in } \Omega_c.
\]

Let $\mathbf{n}_i$ denote the outward unit normal to $\Omega_c$ on
$\Gamma_i$, and define the corresponding tangential vector as
\[
\mathbf{t}_i
=
\begin{bmatrix}
-n_{i,y}\\
n_{i,x}
\end{bmatrix},
\quad i\in\{b,o\}.
\]
Due to the rotational relation between $\widehat{\mathbf{H}}$ and
$\nabla\psi_z$, the measured tangential magnetic field satisfies
\[
H_{t,i}
=
\mathbf{t}_i\cdot\widehat{\mathbf{H}}
=
-\frac{\partial\psi_z}{\partial n_i}.
\]
Hence, the tangential component $H_{t,i}$ naturally provides a Neumann boundary condition for the potential function rather than a Dirichlet condition. The harmonic reconstruction of $\widehat{\mathbf{H}}$ in \eqref{eq:inv_harmonic} is therefore obtained from
$$
\begin{cases}
\Delta\psi_z = 0, & \text{in } \Omega_c, \\
\dfrac{\partial\psi_z}{\partial n_i} = -H_{t,i}, & \text{on } \Gamma_i, \quad i\in\{b,o\}, \\
\int_{\Gamma_b\cup \Gamma_o}\psi_z \, \mathrm{d}l = 0, & \text{(Gauge for Neumann problem)},
\end{cases}
$$
where the last condition fixes the additive constant of $\psi_z$.

Similarly, the magnetic flux density is represented by the harmonic vector
potential $\widehat{\mathbf{A}}$, which under the two-dimensional plane
assumption is written as
\[
\widehat{\mathbf{A}}
=
\begin{bmatrix}
0\\
0\\
\widehat{A}_z(x,y)
\end{bmatrix},
\qquad
\widehat{\mathbf{B}}
=
\nabla\times\widehat{\mathbf{A}}
=
\begin{bmatrix}
\dfrac{\partial \widehat{A}_z}{\partial y}\\[2mm]
-\dfrac{\partial \widehat{A}_z}{\partial x}\\[2mm]
0
\end{bmatrix}.
\]
This representation automatically satisfies the divergence-free condition:
\[
\nabla\cdot\widehat{\mathbf{B}}=0.
\]
Under the local harmonic approximation, the reconstructed flux density is additionally assumed to satisfy
\[
\nabla\times\widehat{\mathbf{B}}=0,
\]
leading to
\[
\Delta \widehat{A}_z=0
\quad
\text{in } \Omega_c.
\]

The measured normal flux density is related to the tangential derivative of the magnetic vector potential by
\[
B_{n,i}
=
\mathbf{n}_i\cdot\widehat{\mathbf{B}}
=
\frac{\partial \widehat{A}_z}{\partial t_i},
\quad
\text{on } \Gamma_i.
\]
Therefore, integrating the measured normal flux density along
each interface determines the boundary trace of $\widehat{A}_z$
up to an interface-dependent additive constant:
\[
\widehat{A}_{z,i}(s)
=
C_i
+
\int_{s_{0,i}}^{s}
B_{n,i}(l)\,\mathrm{d}l,
\qquad i\in\{b,o\},
\]
where $s$ denotes the oriented arc-length coordinate,
$s_{0,i}$ is a reference point on $\Gamma_i$, and
$C_i$ are integration constants.

Then, the harmonic-field approximation of the magnetic flux density in \eqref{eq:inv_harmonic} with Dirichlet boundary conditions is obtained from
\[
\begin{cases}
\Delta \widehat{A}_z = 0,
& \text{in } \Omega_c, \\[1mm]
\widehat{A}_z = \widehat{A}_{z,i},
& \text{on } \Gamma_i, \quad i \in \{b, o\}.
\end{cases}
\]
Consequently, the reconstructed fields satisfy the local harmonic systems described in \eqref{eq:inv_harmonic} with the Cauchy data obtained from \eqref{eq:Cauchy}.
The obtained harmonic fields $\widehat{\mathbf{H}}$ and $\widehat{\mathbf{B}}$ are subsequently used for the field collection and reconstruction of the nonlinear magnetic constitutive characteristics.

\subsection{Data Collection and B--H Curve Reconstruction}

\begin{figure}[htbp]
    \centering
    \hspace{-1cm}
    \includegraphics[width=0.4\textwidth]{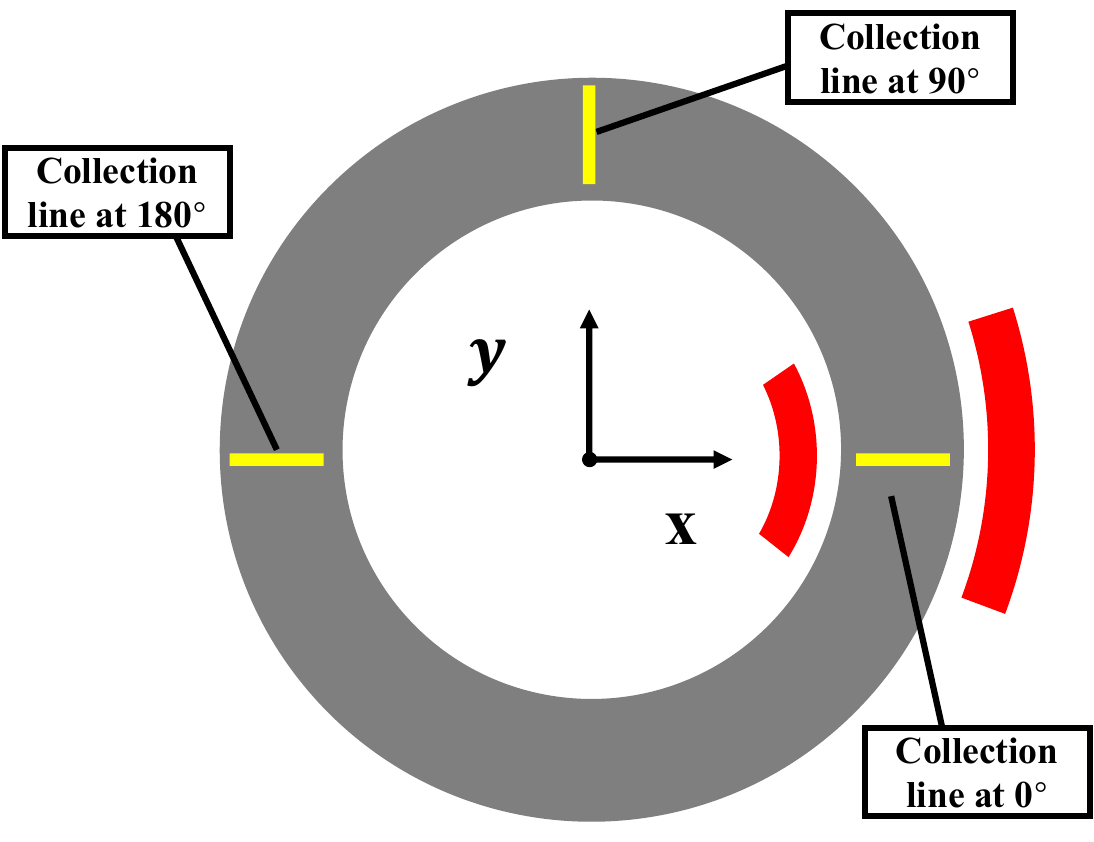}
    \caption{Schematic plot for the field collection at the collection lines.}
    \label{fig:collect}
\end{figure}

To reconstruct the nonlinear magnetic constitutive characteristics, three
collection lines are introduced inside the magnetic core, as illustrated in
Fig.~\ref{fig:collect}. The collection lines, denoted by
$\ell_{0}$, $\ell_{90}$, and $\ell_{180}$, are located at
$0^\circ$, $90^\circ$, and $180^\circ$, respectively.

Due to the geometry of the magnetic core, the magnetic field near these locations is considered to be approximately unidirectional. 
In particular, at the collection lines $\ell_{0}$ and $\ell_{180}$, the magnetic field is predominantly oriented along the $y$-direction, whereas at $\ell_{90}$ it is predominantly oriented along the $x$-direction.
Therefore, the field components are approximated as
\[
\begin{aligned}
&\widehat{H}_x \approx 0,\quad \widehat{B}_x \approx 0,
&&\text{on } \ell_{0}\cup\ell_{180},\\
&\widehat{H}_y \approx 0,\quad \widehat{B}_y \approx 0,
&&\text{on } \ell_{90}.
\end{aligned}
\]
Consequently, the data collected at $\ell_{0}$ and $\ell_{180}$ provide
samples of the $y$-direction magnetic characteristic, while the data at
$\ell_{90}$ provide samples of the $x$-direction characteristic.

To reduce the influence of local field variations, the reconstructed fields are spatially averaged along each collection line.
 For a collection line $\ell_{\alpha}$, with $\alpha\in\{0,90,180\}$, the averaged field components for $q\in\{x,y\}$ are defined as
\begin{equation}
\overline{H}_{q,\alpha}
=
\frac{1}{|\ell_{\alpha}|}
\int_{\ell_{\alpha}} \widehat{H}_q\,\mathrm{d}l,
\quad
\overline{B}_{q,\alpha}
=
\frac{1}{|\ell_{\alpha}|}
\int_{\ell_{\alpha}} \widehat{B}_q\,\mathrm{d}l,
\label{eq:average}
\end{equation}
where $|\ell_{\alpha}|$ denotes the length of the corresponding collection
line.

Accordingly, the effective magnetic data used for the reconstruction are
\[
\left(
\overline{H}_{y,0},
\overline{B}_{y,0}
\right),
\quad
\left(
\overline{H}_{x,90},
\overline{B}_{x,90}
\right),
\quad
\left(
\overline{H}_{y,180},
\overline{B}_{y,180}
\right).
\]

By applying a sequence of different source-current excitations
$\mathbf{J}^{(k)}$, $k=1,\ldots,N_J$, different magnetic operating points
are generated. The corresponding averaged field pairs are collected as
\[
\begin{aligned}
\mathcal{D}_x
&=
\left\{
\left(
\overline{H}_{x,90}^{(k)},
\overline{B}_{x,90}^{(k)}
\right)
\right\}_{k=1}^{N_J},\\
\mathcal{D}_y
&=
\left\{
\left(
\overline{H}_{y,0}^{(k)},
\overline{B}_{y,0}^{(k)}
\right),
\left(
\overline{H}_{y,180}^{(k)},
\overline{B}_{y,180}^{(k)}
\right)
\right\}_{k=1}^{N_J}.
\end{aligned}
\]
These data sets provide discrete samples of the nonlinear magnetic
characteristics in the two principal directions and are subsequently used
to reconstruct
\[
\widehat{B}_x=\mathcal{B}_x(\widehat{H}_x),
\quad
\widehat{B}_y=\mathcal{B}_y(\widehat{H}_y).
\]

\subsection{Quantitative Comparison and Error Metrics}

For each excitation $k$, the reconstructed magnetic field strength is mapped
through the prescribed constitutive law to obtain the reference flux density.
For a total of $K$ excitation currents, the reference value and mean relative
error are defined compactly as
\begin{equation}
\begin{aligned}
B_{q,\mathrm{ref}}^{(k)}
&=\mathcal{B}_{q,\mathrm{ref}}\!\left(\widehat{H}_q^{(k)}\right),\\
E_q
&=\frac{100\%}{K}\sum_{k=1}^{K}
\frac{\left|\widehat{B}_q^{(k)}-B_{q,\mathrm{ref}}^{(k)}\right|}
{\left|B_{q,\mathrm{ref}}^{(k)}\right|},
\qquad q\in\{x,y\}.
\end{aligned}
\end{equation}
Here, $\widehat{H}_q^{(k)}$ and $\widehat{B}_q^{(k)}$ denote the averaged
reconstructed fields at the corresponding collection line.
The B--H characteristic of NO27 electrical steel, denoted by $\mathcal{B}_{\mathrm{NO27}}$, is used as the reference curve \cite{daemModelingInterlocking2020}.
The reference B--H curve is assigned to the $y$-direction,
corresponding to the rolling direction (RD) and representing
the easier magnetization direction. The $x$-direction corresponds
to the transverse direction (TD), for which the magnetic field
strength required to reach the same flux density is increased
by a factor of three.

Accordingly, the two principal reference magnetization characteristics are defined as
\[
\begin{aligned}
\mathcal{B}_{x,\mathrm{ref}}(H_x) &:= \mathcal{B}_{\mathrm{NO27}}\!\left(\frac{H_x}{3}\right),
\\
\mathcal{B}_{y,\mathrm{ref}}(H_y) &:= \mathcal{B}_{\mathrm{NO27}}(H_y).
\end{aligned}
\]

This scaling provides an easily controlled anisotropic test case for evaluating the proposed identification method.

\section{Numerical Simulations and Reconstruction Results}
\subsection{Reconstruction from Unperturbed Data}

\begin{figure*}[t]
    \centering
    \begin{subfigure}[t]{0.31\textwidth}
        \centering
        \includegraphics[width=\textwidth]{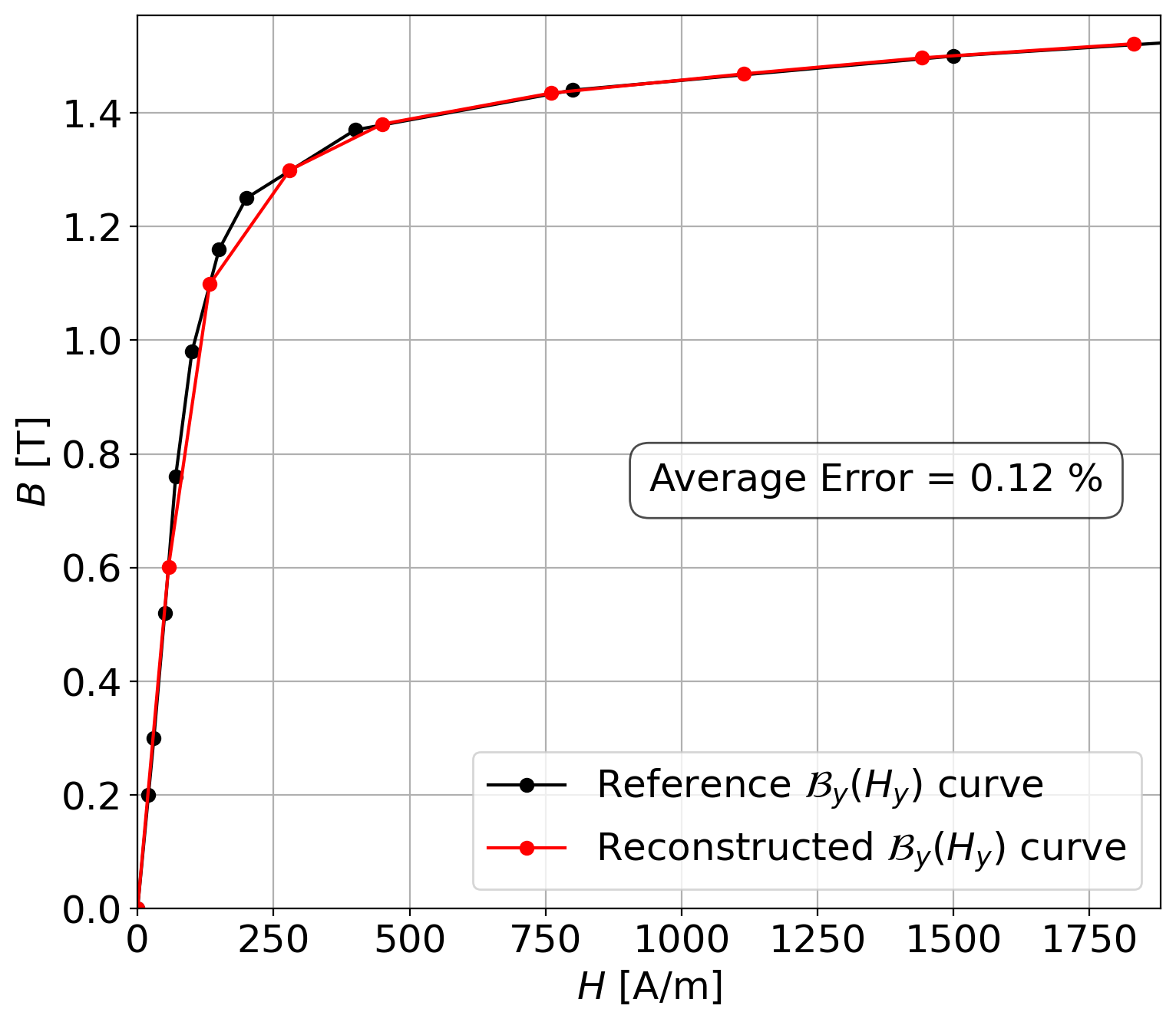}
        \caption{$0^\circ$ collection line.}
        \label{fig:BH_0}
    \end{subfigure}
    \hfill
    \begin{subfigure}[t]{0.31\textwidth}
        \centering
        \includegraphics[width=\textwidth]{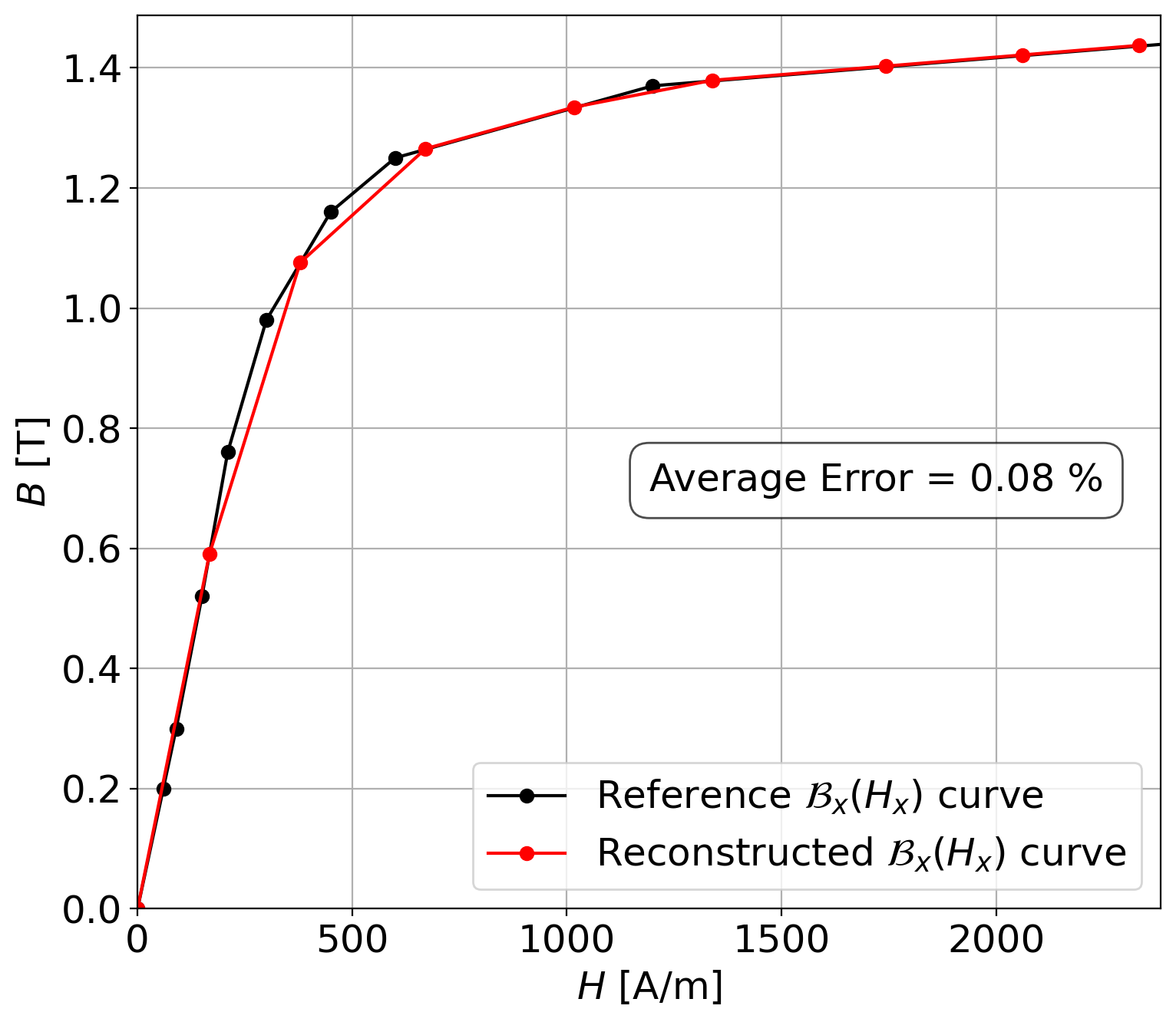}
        \caption{$90^\circ$ collection line.}
        \label{fig:BH_90}
    \end{subfigure}
    \hfill
    \begin{subfigure}[t]{0.31\textwidth}
        \centering
        \includegraphics[width=\textwidth]{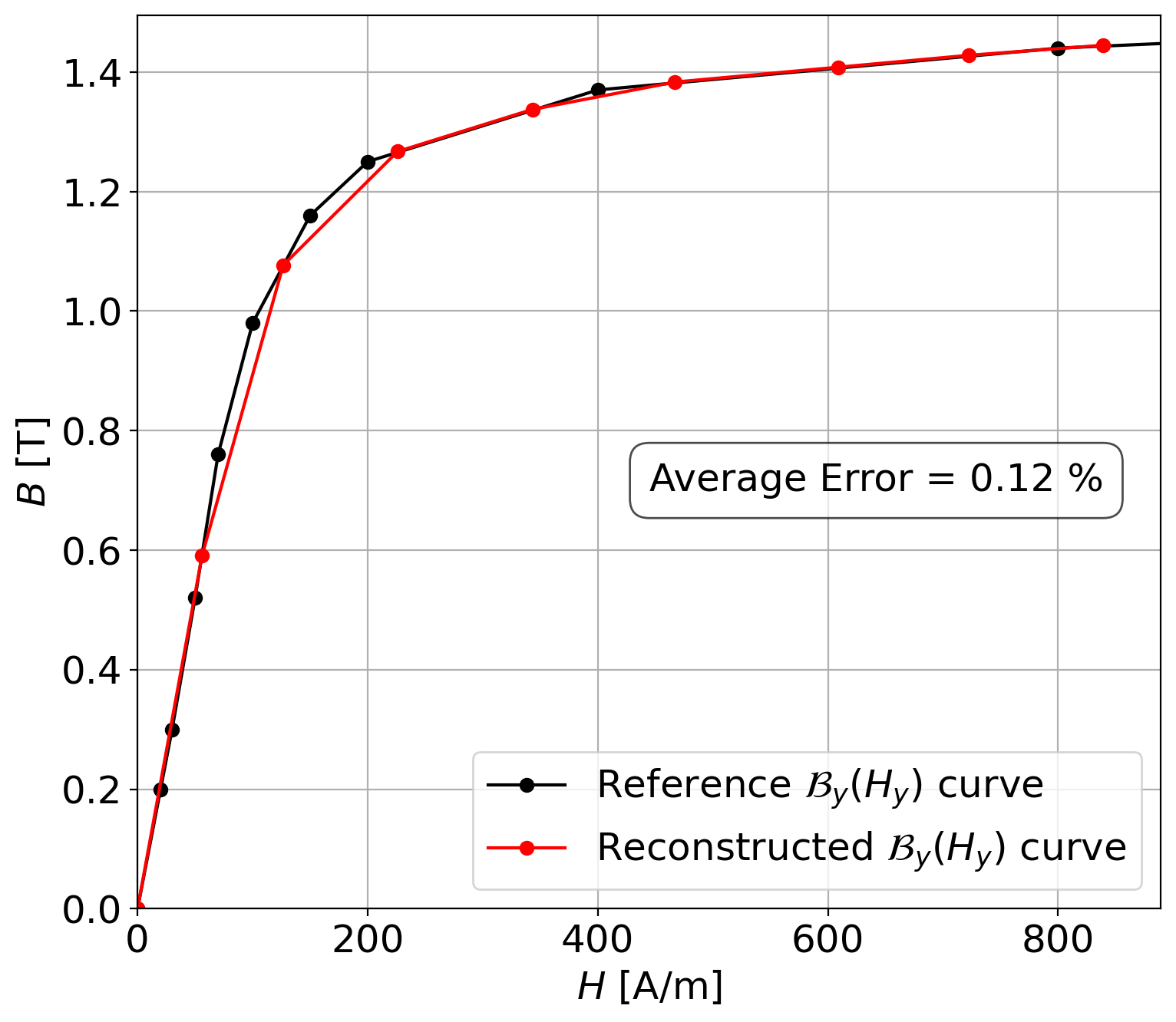}
        \caption{$180^\circ$ collection line.}
        \label{fig:BH_180}
    \end{subfigure}

    \caption{Reconstructed magnetic constitutive characteristics obtained from the three collection lines.}
    \label{fig:BH_reconstruction}
\end{figure*}

The method is first evaluated using unperturbed Cauchy data obtained from the
forward solution. As shown in Fig.~\ref{fig:BH_reconstruction}, the reconstructed
material properties curves at all three collection lines closely follow the prescribed
B--H curves, capturing both the nonlinear low-field region and the gradual
saturation at higher field strengths. The corresponding mean relative errors
are summarized in Table~\ref{tab:unperturbed_error}.
\begin{table}[htbp]
    \centering
    \caption{Mean relative reconstruction errors for the unperturbed data.}
    \label{tab:unperturbed_error}
    \begin{tabular}{ccc}
        \hline
        Collection line & Reconstructed characteristic & Mean relative error \\
        \hline
        $0^\circ$   & $\mathcal{B}_y(H_y)$ & $0.12\%$ \\
        $90^\circ$  & $\mathcal{B}_x(H_x)$ & $0.08\%$ \\
        $180^\circ$ & $\mathcal{B}_y(H_y)$ & $0.12\%$ \\
        \hline
    \end{tabular}
\end{table}

The maximum mean relative error is $0.12\%$, while the $90^\circ$ collection
line gives the minimum error of $0.08\%$. This confirms the accuracy of the
harmonic reconstruction for unperturbed boundary data.

\begin{figure*}[t]
    \centering
    \begin{subfigure}[t]{0.31\textwidth}
        \centering
        \includegraphics[width=\textwidth]{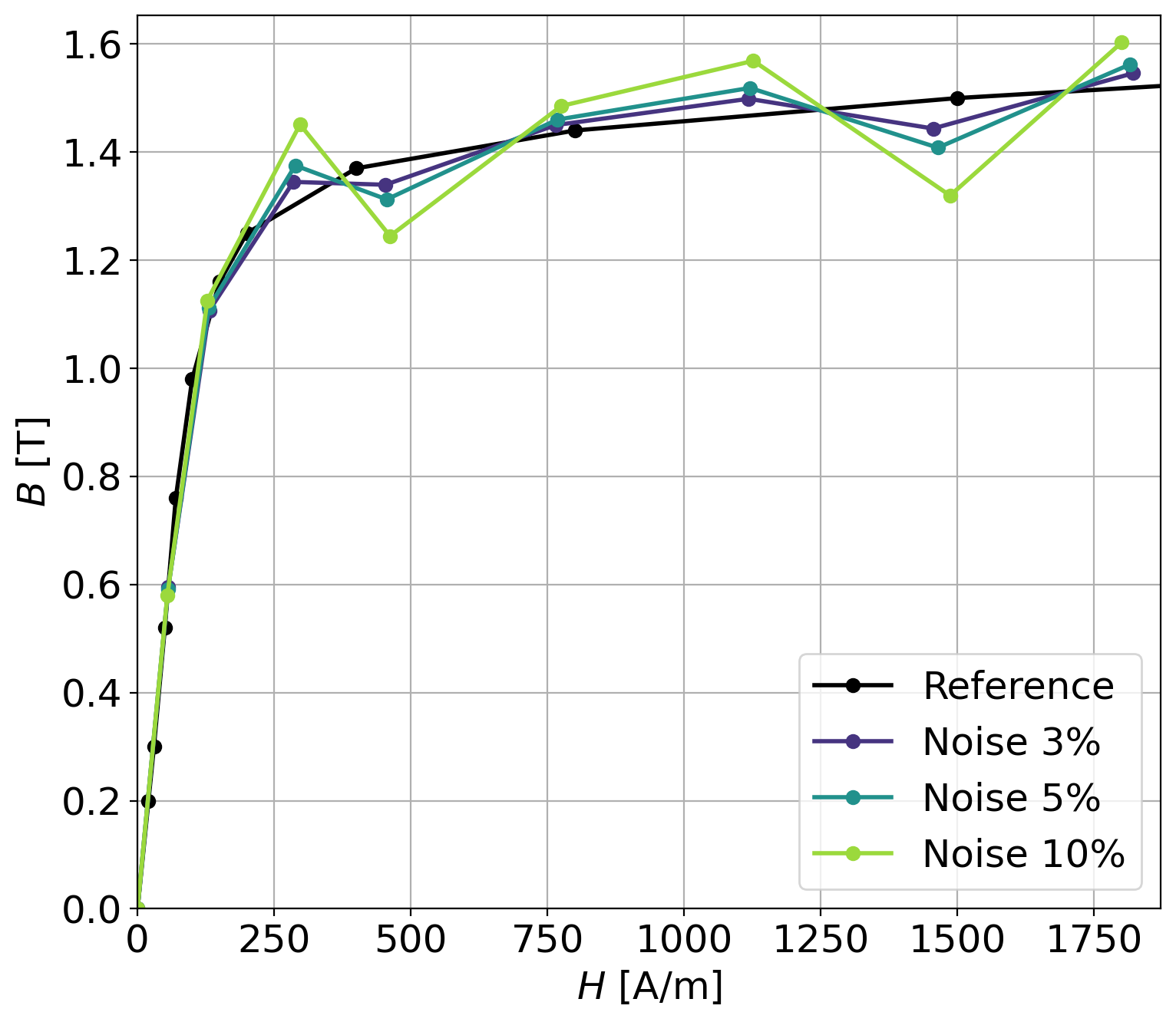}
        \caption{$0^\circ$ collection line.}
        \label{fig:noise_comparison_0}
    \end{subfigure}
    \hfill
    \begin{subfigure}[t]{0.31\textwidth}
        \centering
        \includegraphics[width=\textwidth]{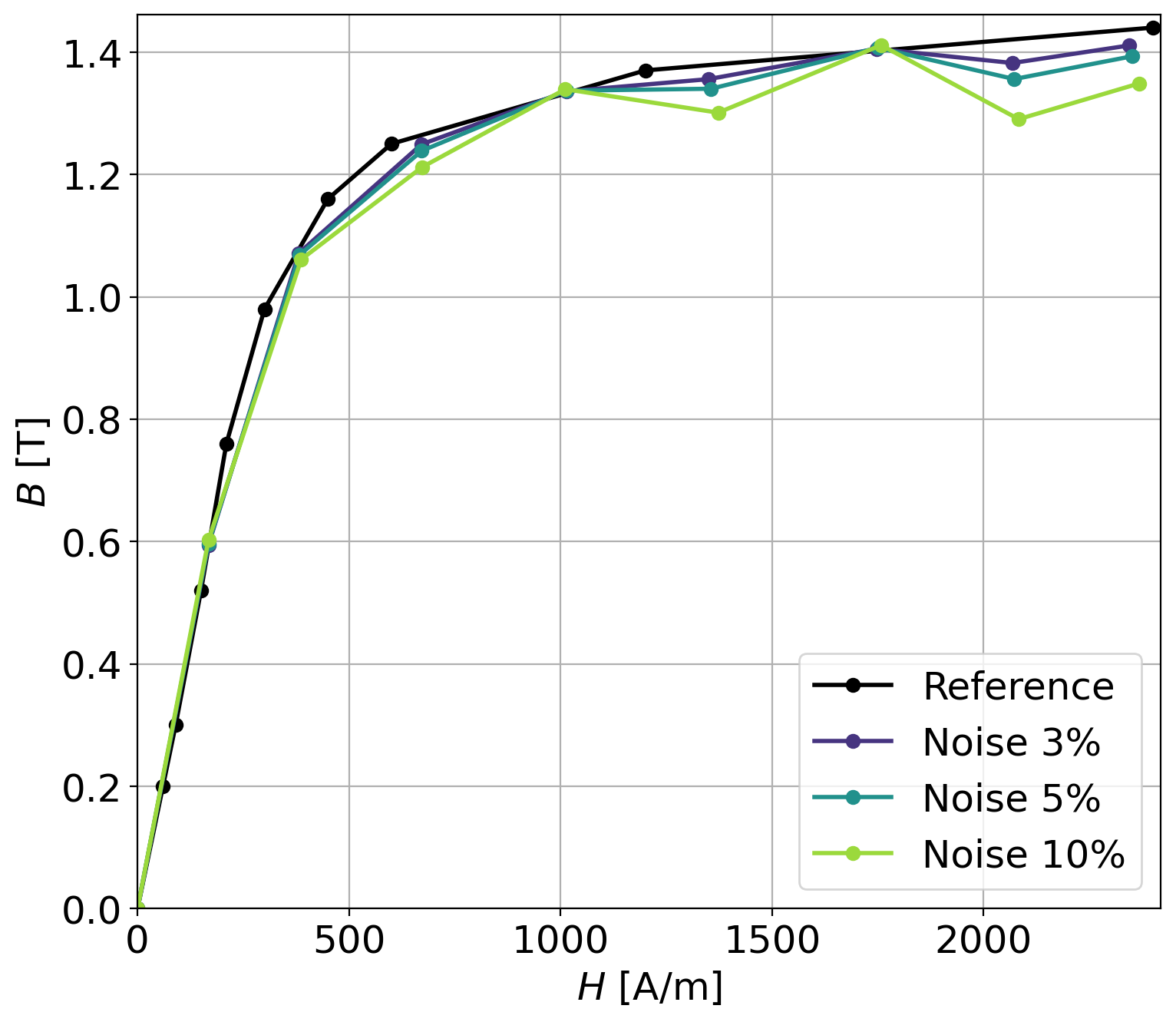}
        \caption{$90^\circ$ collection line.}
        \label{fig:noise_comparison_90}
    \end{subfigure}
    \hfill
    \begin{subfigure}[t]{0.31\textwidth}
        \centering
        \includegraphics[width=\textwidth]{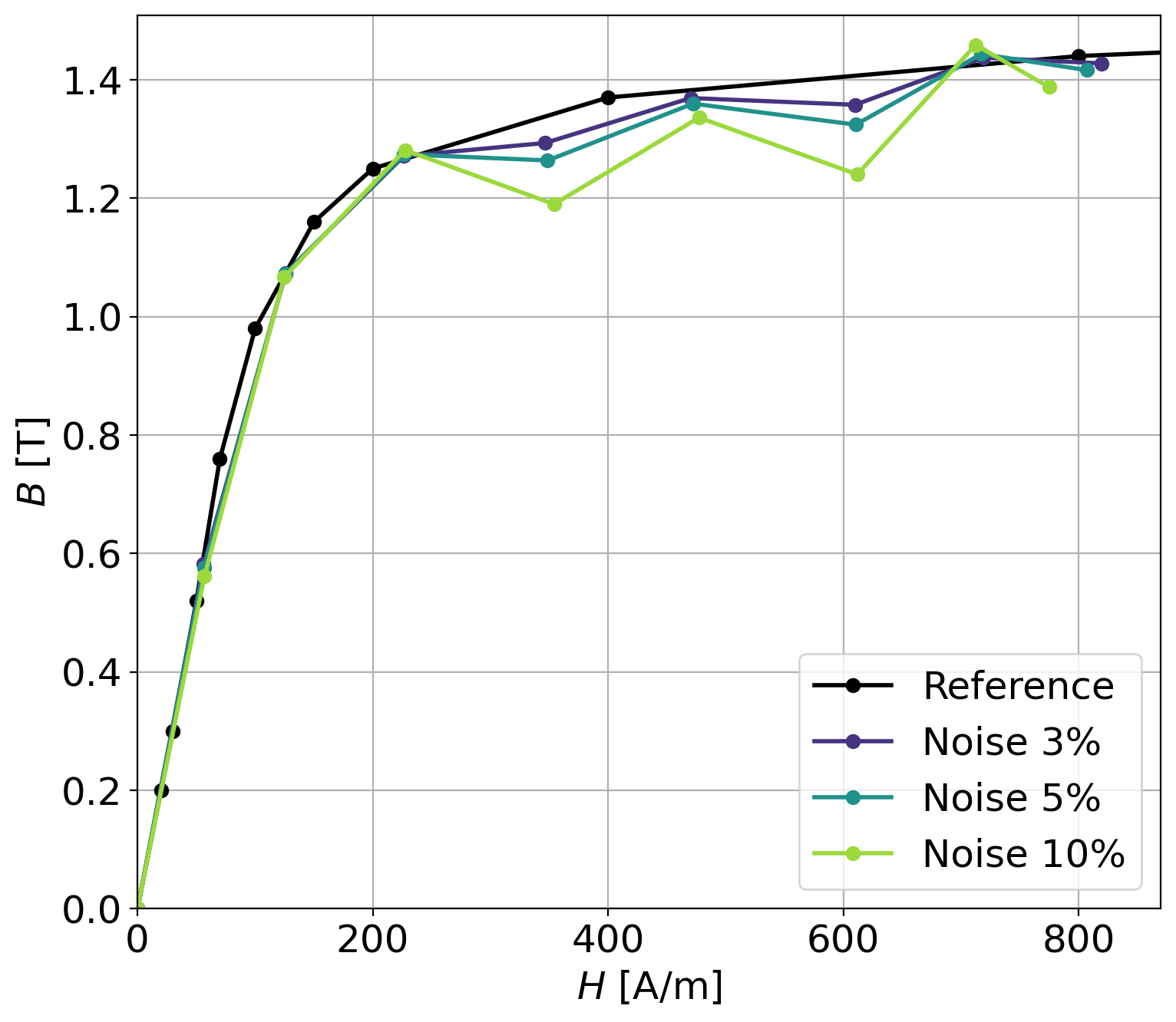}
        \caption{$180^\circ$ collection line.}
        \label{fig:noise_comparison_180}
    \end{subfigure}
    \caption{Reconstructed magnetic characteristics under different noise levels.}
    \label{fig:noise_comparison}
\end{figure*}

\subsection{Reconstruction from Perturbed Data and Robustness Analysis}

To evaluate the robustness of the proposed reconstruction method against measurement uncertainties, additive Gaussian noise is introduced directly into the Cauchy data.
 For a generic boundary quantity $g$, the perturbed data are defined as
\[
\widetilde{g}_j
=
g_j+\varepsilon_j,
\qquad
\varepsilon_j
\sim
\mathcal{N}(0,\sigma_g^2),
\]
where
\[
\sigma_g
=
\eta\,g_{\mathrm{RMS}},
\qquad
g_{\mathrm{RMS}}
=
\sqrt{
\frac{1}{N}
\sum_{j=1}^{N}
|g_j|^2
}.
\]
Here, $\eta$ denotes the prescribed noise level.
In this study, $\eta=3\%$, $5\%$, and $10\%$ are considered.
The noise is independently generated for the Cauchy data associated with $B_n$ and $H_t$ on both interfaces. It represents random uncertainty from the Hall probe and data-acquisition system in the measurement procedure described in Section~\ref{section:measurement}; systematic effects such as probe misalignment and finite stand-off distance are not included.

Figure~\ref{fig:noise_comparison} shows that the deviations increase with the
noise level, while the overall nonlinear trend remains identifiable in all
three collection lines even at $10\%$ noise. The corresponding mean relative errors
in the reconstructed flux density are reported in Table~\ref{tab:noise_error}.

\begin{table}[htbp]
    \centering
    \caption{Mean relative $B$ reconstruction errors under different noise levels.}
    \label{tab:noise_error}
    \begin{tabular}{ccccc}
        \hline
        Noise level
        & $0^\circ$
        & $90^\circ$
        & $180^\circ$
        & All collection lines \\
        \hline
        $3\%$  & $1.80\%$ & $1.01\%$ & $1.36\%$ & $1.39\%$ \\
        $5\%$  & $2.99\%$ & $1.69\%$ & $2.26\%$ & $2.32\%$ \\
        $10\%$ & $5.99\%$ & $3.40\%$ & $4.50\%$ & $4.63\%$ \\
        \hline
    \end{tabular}
\end{table}

The overall mean error increases from $1.39\%$ at $3\%$ noise to $2.32\%$ and
$4.63\%$ at $5\%$ and $10\%$ noise, respectively. The $90^\circ$ collection
line, which corresponds to the harder transverse direction, consistently gives
the smallest error. These results indicate that the
harmonic reconstruction remains stable under moderate perturbations of the
Cauchy data.

\section{Discussion}

The main capabilities of the proposed identification framework can be summarized as follows:
\begin{itemize}
    \item \textbf{Model-free identification:}
    the inverse procedure does not require a predefined analytical model for the nonlinear magnetic constitutive law. The internal magnetic characteristics are reconstructed directly from the measured Cauchy data.

    \item \textbf{Simplified inverse formulation:}
    instead of iteratively solving a nonlinear constitutive problem, the inverse problem is reduced to linear harmonic reconstruction problems for $\mathbf{B}$ and $\mathbf{H}$, which significantly simplifies the numerical implementation.

    \item \textbf{Robustness against measurement noise:}
    the numerical results show that the reconstructed $B$--$H$ characteristics remain close to the reference curves under moderate perturbations of the Cauchy data. Even with a $10\%$ noise level, the overall mean reconstruction error remains below $5\%$.
\end{itemize}

The main limitations are:
\begin{itemize}
    \item \textbf{Dependence on the harmonic approximation:}
    while $\nabla\times\mathbf{H}=0$ and $\nabla\cdot\mathbf{B}=0$ follow directly from the magnetostatic equations inside the current-free core, the additional assumptions
    \[
    \nabla\cdot\widehat{\mathbf{H}}=0,
    \qquad
    \nabla\times\widehat{\mathbf{B}}=0
    \]
    are introduced specifically for the harmonic reconstruction. The error may increase when the material is highly non-uniform, anisotropic, or geometrically complex.

    \item \textbf{Restriction to single-valued magnetization characteristics:}
    the present study considers only single-valued $B$--$H$ curves. Demagnetization behavior and hysteresis effects are outside the scope of this work.

    \item \textbf{Requirement of boundary measurements:}
    the method requires the normal component of $\mathbf{B}$ and the tangential component of $\mathbf{H}$ on both core interfaces. In practice, the Hall-probe measurements may be affected by finite probe size and stand-off distance, positioning and orientation errors, limited access near the winding, sensor calibration, and temperature drift. These effects can introduce correlated or systematic errors beyond the independent Gaussian perturbations considered here and should be assessed in future experimental validation.
\end{itemize}

\section{Conclusion}
This work presents a model-free approach for identifying nonlinear anisotropic magnetic constitutive characteristics from boundary Cauchy data.
By independently reconstructing the magnetic flux density and magnetic field strength using harmonic approximations, the proposed inverse formulation avoids the direct use of the unknown constitutive law.
The reconstructed fields are subsequently sampled along selected collection lines to recover the principal-direction $B$--$H$ characteristics.
Numerical results demonstrate high reconstruction accuracy for unperturbed data and good robustness under measurement noise.
The method therefore provides a simple and computationally efficient framework for magnetic constitutive law identification, while its accuracy remains dependent on the validity of the harmonic and locally unidirectional field approximations.

\bibliographystyle{IEEEtran}
\bibliography{ref}

\vspace{12pt}
\color{red}
\end{document}